\documentclass[a4paper,11pt]{article}

\usepackage{xcolor}

\newcommand{\boxnumber}[1]{%
  \fcolorbox{black}{white}{\raisebox{-.1ex}{\small #1}}%
}

\usepackage{fourier}
\usepackage{color}
\usepackage{graphicx}
\usepackage{url}
\usepackage[affil-it]{authblk}
\usepackage{amsmath}
\usepackage{wrapfig}
\usepackage{bm}
\usepackage[T1]{fontenc}
\usepackage{times}
\usepackage{multirow}
\usepackage{etoolbox}
\usepackage{caption}
\usepackage{marvosym}

\DeclareMathAlphabet{\mathcal}{OMS}{cmsy}{m}{n}

\AtBeginEnvironment{tabular}{\footnotesize}

\begin{document}

\title{Diffusion Model-based Data Augmentation Method\\ for Fetal Head Ultrasound Segmentation}

\author[1,2]{Fangyijie Wang \thanks{Corresponding author: fangyijie.wang@ucdconnect.ie}}

\author[3]{Kevin Whelan}
\author[3]{F\'elix Balado}
\author[1,2]{Kathleen M. Curran}
\author[1,3]{Gu\'enol\'e Silvestre}

\affil[1]{Research Ireland Centre for Research Training in Machine Learning}
\affil[2]{School of Medicine, University College Dublin, Dublin, Ireland}
\affil[3]{School of Computer Science, University College Dublin, Dublin, Ireland}
\date{}
\maketitle
\thispagestyle{empty}

\begin{abstract}
Medical image data is less accessible than in other domains due to privacy and regulatory constraints. In addition, labeling requires costly, time-intensive manual image annotation by clinical experts. To overcome these challenges, synthetic medical data generation offers a promising solution. Generative AI (GenAI), employing generative deep learning models, has proven effective at producing realistic synthetic images.
This study proposes a novel mask-guided GenAI approach using diffusion models to generate synthetic fetal head ultrasound images paired with segmentation masks. These synthetic pairs augment real datasets for supervised fine-tuning of the Segment Anything Model (SAM).
Our results show that the synthetic data captures real image features effectively, and this approach reaches state-of-the-art fetal head segmentation, especially when trained with a limited number of real image-mask pairs. In particular, the segmentation reaches Dice Scores of 94.66\% and 94.38\% using a handful of ultrasound images from the Spanish and African cohorts, respectively. Our code, models, and data are available on 
\url{https://github.com/13204942/Diffusion-based_DA_Fetal_Ultrasound_Segmentation}.

\end{abstract}

\section{Introduction}
Deep learning has shown great potential in performing various medical image analysis tasks such as classification, segmentation, and object detection across multiple imaging modalities (Computed Tomography, X-ray, and ultrasound) \cite{Zhou:2021}. One of the challenges in applying deep learning to ultrasound imaging tasks is the scarcity of large labeled datasets for supervised model training. There are several reasons for this. Firstly, there are legal, privacy (e.g., de-identification risk), and security concerns around sharing medical data outside of the medical institution where the data is generated \cite{Piaggio:2021}. In addition, for medical imaging analysis, significant time and effort from an expert is required to annotate the images for the task of interest, e.g., fetal head segmentation for gestational age estimation \cite{Salomon:2011,Sarris:2012}, creating an annotation mask around head skull areas of interest for segmentation tasks \cite{Heuvel:2018_b,poojari:2022}. 
To tackle these problems, researchers looked to Generative AI (GenAI) utilizing generative deep learning models to create synthetic medical images to fill the lack of data availability gap \cite{Dhariwal:2021,Sastry:2024}.
In medical image analysis, these models can be trained with a relatively small number of real medical images and used to generate an unlimited number of synthetic images, similar to the training set \cite{Kazerouni:2023}. 

Recent studies demonstrate the effectiveness of GenAI in diverse fetal ultrasound image synthesis applications \cite{Lee:2020,Montero:2021,Whelan:2024,Lasala:2024,Duan:2025,Wang:2025}.
\cite{Lee:2020,Lasala:2024} developed methods to generate synthetic fetal brain ultrasound images using Generative Adversarial Network (GAN) model. \cite{Montero:2021} improved fetal head ultrasound plane classification with synthetic images generated by GAN. \cite{Duan:2025,Wang:2025} demonstrated the diffusion model-based method for synthetic fetal ultrasound image generation in plane classification tasks. 
\cite{Lasala:2024} introduces an approach that leverages class activation maps (CAM) as a prior condition to generate standard planes of the fetal head using a conditional GAN model. 
However, these studies only investigate the effectiveness of synthetic fetal head ultrasound images in simple classification tasks. The GenAI models utilized in these studies were trained with many real-world images. \cite{Whelan:2024} proposes an in-channel mask injection method to generate synthetic fetal ultrasound images using a diffusion model. Unlike their approach and experiments, we comprehensively analyze fetal head segmentation performance using synthetic images on diverse datasets.

This paper investigates techniques to jointly generate synthetic fetal ultrasound images and their corresponding fetal head segmentation mask using Stable Diffusion (SD). We first fine-tune SD using the Low-Rank Adaptation (LoRA) method using a small number of real fetal ultrasound images. Then, we propose an in-channel mask injection method to incorporate the mask generation with the image generation. This method replaces one of the three RGB channels of the input image with the segmentation mask from the training examples to provide input to the SD model. 
Subsequently, we evaluate the efficacy of synthetic images in semantic segmentation using a foundational model known as the Segment Anything Model (SAM).
The key contributions of this paper are four-fold: 
1) We propose a novel mask-guided diffusion model for generating synthetic images in the ultrasound domain. 2) We demonstrate an efficient and effective diffusion model-based data augmentation with a few-shot learning strategy. 3) Our data augmentation technique significantly improves the generalization of SAM in fetal ultrasound image segmentation across three diverse datasets. 4) The results illustrate that our approach achieves state-of-the-art (SOTA) performance when the number of annotated images is fewer than 50.
\section{Methods}

\subsection{Preliminaries}

This section reviews the key models and techniques that form the foundation of our approach, including Stable Diffusion, LoRA fine-tuning, and the Segment Anything Model (SAM).

\paragraph{Stable Diffusion (SD):} The SD model is a SOTA text-to-image generation model developed by Stability AI \cite{Rombach:2022}. The model begins with an initial noisy image and iteratively refines it through denoising steps guided by the provided text prompt or other conditioning signals. This process leverages a reverse diffusion process trained on large-scale image-text datasets. The architecture involves U-Net and Transformer components to capture spatial and contextual information effectively. The pre-trained SD model does not generate realistic fetal ultrasound images and must be fine-tuned with HC18 samples. The fine-tuning method we employ is LoRA.

\noindent {\bf LoRA:} LoRA \cite{Hu:2022} is a fine-tuning technique designed to optimize the training of large-scale language models by introducing a low-rank decomposition approach. Traditional fine-tuning methods update all the parameters of a pre-trained model, which can be computationally expensive and memory-intensive. LoRA addresses these challenges by injecting trainable low-rank matrices into each layer of the Transformer architecture, significantly reducing the number of trainable parameters.
Specifically, let $W_0 \in \mathbb{R}^{d \times k}$ represent the weight matrix of a given layer in the pre-trained model. Instead of updating $W_0$ directly, LoRA decomposes the weight update into two smaller matrices $A \in \mathbb{R}^{d \times r}$ and $B \in \mathbb{R}^{r \times k}$, where $r \ll \text{min}(d,k)$ is the matrix rank. The weight update can be expressed as: $\Delta W=A B$. During fine-tuning, the modified weight matrix $W$ becomes: $W=W_0+\Delta W=W_0+A B$. By constraining $A$ and $B$ to be of low rank, LoRA reduces the number of trainable parameters from $d \times k$ to $r \times (d + k)$.

\noindent{\bf Segment Anything Model (SAM):} SAM is a segmentation model developed by Meta 
AI~\cite{kirillov:2023}. Its primary objective is to simplify the segmentation process by serving as a foundational model that can be prompted with various inputs, including clicks, boxes, or text, making it accessible to a broad range of downstream applications. The SAM model $f_\text{SAM}$ consists of three main components: an image encoder based $\mathcal{E}_\text{SAM}$ on a Masked Autoencoder (MAE) pre-trained Vision Transformer (ViT) \cite{dosovitskiy:2021}, a prompt encoder $\mathcal{E}_\text{Prompt}$ based on the text-encoder of the CLIP \cite{radford:2021} model and a mask decoder $\mathcal{D}_\text{SAM}$ which maps the image embedding, prompt embedding and an output token to a mask. 

\begin{figure}[!hb]
    \centering
    \includegraphics[width=\textwidth]{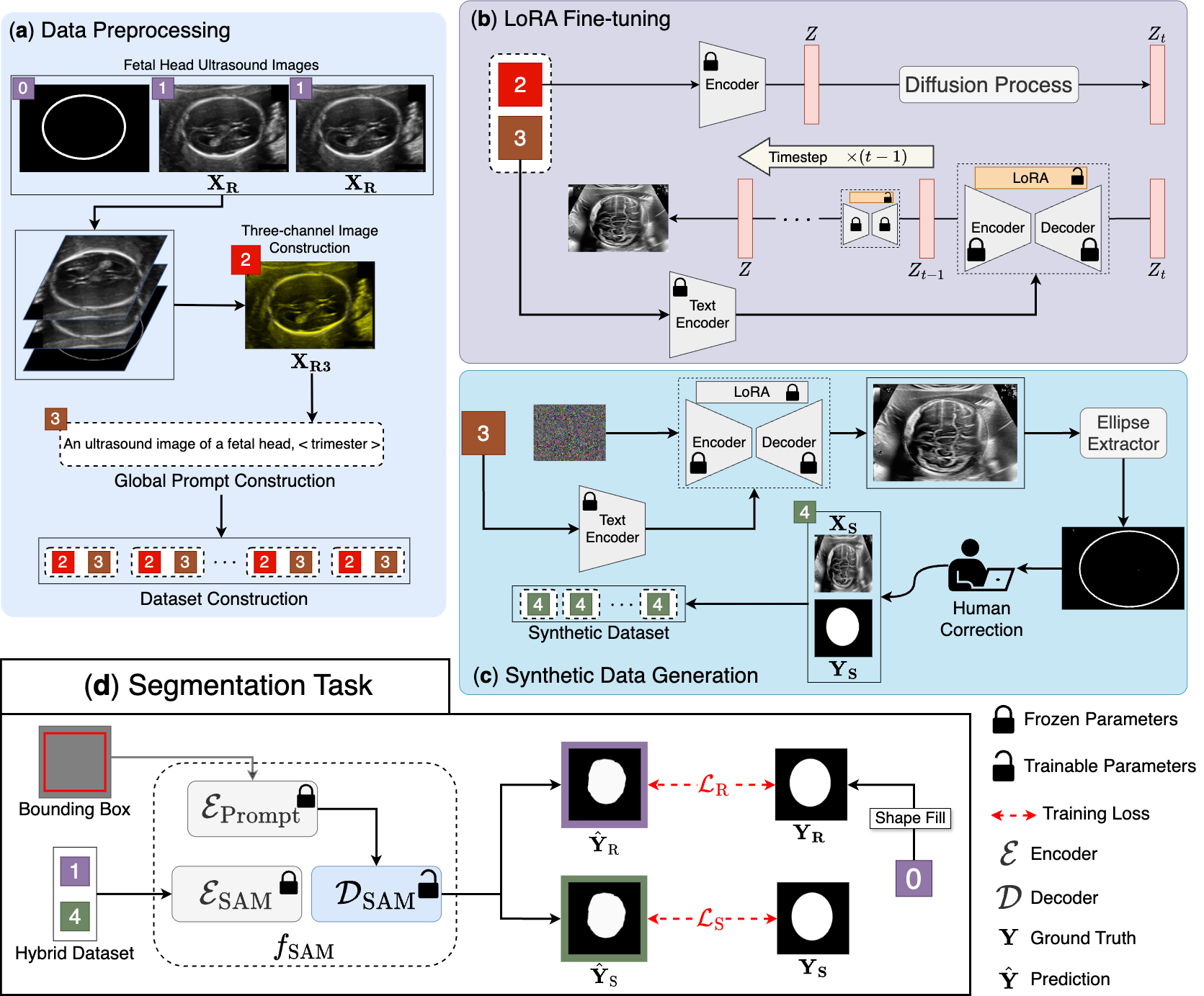}
    \caption{The overview of our proposed diffusion-based data augmentation method for fetal ultrasound images. It consists of three main steps: ({\bf a}) Concatenating the gray-scale ultrasound image $\boldsymbol{X}_\text{R}$ and the segmentation mask to obtain a three-channel image $\boldsymbol{X}_\text{R3}$. ({\bf b}) Adopting the LoRA technique to fine-tune the SD model with real images and text prompts. ({\bf c}) A fine-tuned SD model generates synthetic images solely with text prompts. ({\bf d}) Constructing a hybrid dataset with synthetic images to improve fetal ultrasound segmentation performance of a pre-trained SAM $f_\text{SAM}$.}
    \vspace{-0pt}
    \label{fig:our_methods}    
\end{figure}

\vspace*{-1em}
\subsection{Few-shot Fine-tuning Diffusion Model}

Our approach follows two main steps: \textbf{Data Preprocessing} and \textbf{LoRA Fine-tuning}, as depicted in Figure~\ref{fig:our_methods}. In the Data Preprocessing stage, a three-channel image is constructed with an in-channel mask, as shown in Figure~\ref{fig:our_methods}(a). This three-channel image $\boldsymbol{X}_\text{R3}$ is used to fine-tune Stable Diffusion with LoRA. Through fine-tuning, SD learns to generate three-channel images where one of the channels contains an ellipse-like grayscale mask and the other two channels contain synthetic grayscale ultrasound images. This approach produces precise segmentation masks and realistic images, making it the preferred method for generating synthetic data to fine-tune SAM.

In addition, SD uses the text prompt as input to control the quality of the final generated image. As the HC18 dataset contains images from three trimesters, we experimented with using a prompt for each trimester during fine-tuning. These prompts are created based on a template, {\it "An ultrasound image of a fetal head, <trimester>"} where {\it "<trimester>"} is replaced with the specific trimester period, such as "first trimester", "second trimester", and "third trimester". The prompts are used to construct a dataset for training SD; refer to dataset \boxnumber{3} in Figure~\ref{fig:our_methods}(a). We can generate images from different trimesters by supplying a trimester-specific prompt during inference. Because we aim to employ a few-shot fine-tuning strategy to let the model learn the relationship between the prompt and images more effectively, we improved the realism of generated images by fine-tuning three separate LoRA models, one for each trimester.

In the LoRA Fine-tuning stage, we employ a few-shot learning strategy to fine-tune SD, requiring only a few sample images (from 20 to 25) to generate high-quality synthetic images. In Figure~\ref{fig:our_methods}(b), in denoising latent space $Z_t$ during $t$ timestep, the LoRA technique updates some weights of the U-Net \cite{Ronneberger:2015} within the diffusion model and keeps the whole U-Net frozen. This technique significantly reduces the computational resources required and the time costs.

\subsection{Synthetic Image Generation}
In Figure~\ref{fig:our_methods}(c), we demonstrate the usage of text prompts from \boxnumber{3} to generate synthetic ultrasound images. During the generation, the fine-tuned SD with LoRA weights is used for inference. We use the same text encoder to encode text prompts into the latent space. After the generation process, we implemented an ellipse extraction program in Python to obtain the contour of ellipses, the Ellipse Extractor. This program utilizes a threshold value for image thresholding using the OpenCV library \cite{opencvlib} to identify contours. Nevertheless, residual noisy pixels persist, so we do manual correction by human operators.
Furthermore, we manually reviewed each ellipse mask and filled the ellipse shape to get the ground truth binary mask $\boldsymbol Y_{S}$ in dataset \boxnumber{4} as shown in Figure~\ref{fig:our_methods}(c).

\subsection{Fine-tuning SAM}

The SAM can utilize different sizes of ViTs in the image encoder: base (ViT-B), large (ViT-L), and huge (ViT-H). We use the ViT-B model as the image encoder as done in \cite{Ma:2024}. During fine-tuning, the parameters of the image encoder $\mathcal{E}_\text{SAM}$ and prompt encoder $\mathcal{D}_\text{Prompt}$ are frozen as shown Figure~\ref{fig:our_methods}(d). Only the mask decoder $\mathcal{D}_\text{SAM}$ parameters are updated during training. Therefore, we make it practical to fine-tune the model using a single GPU. 
As all models evaluated are variants of SAM, they require an input prompt. We provide a bounding box prompt, which is the bounding box of the ground truth segmentation mask $\boldsymbol{Y}$ with a random perturbation of $[0,20]$ pixels to simulate a manual annotation, as done in \cite{Ma:2024}. In Figure~\ref{fig:our_methods}(d), given an input image as $\boldsymbol{X} \in \mathbb{R}^{H \times W \times 3}$ from a hybrid dataset \{\boxnumber{1}\,,\,\boxnumber{4}\} and a ground truth mask $\boldsymbol{Y} \in \mathbb{R}^{H \times W \times 1}$ from dataset \boxnumber{0}. The input visual prompt for SAM is denoted as $\boldsymbol{P} \in \mathbb{R}^{H \times W \times 3}$, where $H \times W$ are the spatial dimensions. In this study, $\boldsymbol{P}$ is denoted as a bounding box $\boldsymbol{P}_\text{box} \in \mathbb{R}^{(H-q) \times (H-q) \times 3}$, where $q$ is the number of random perturbations of pixels in the range $[0,20]$.

\section{Experiments}


\paragraph{Data Description:}

We utilize three public datasets in this study: the {\bf HC18} dataset is collected from the database in Netherlands~\cite{Heuvel:2018_b}; the Spanish transthalamic dataset ({\bf ES}) is acquired from two centers in Spain~\cite{Xavier:2020}; the {\bf AF} dataset contains images collected from five African countries \cite{balcells_data:2023,sendra_balcells:2023}. 
The {\bf HC18} dataset contains 999 fetal ultrasound images without growth abnormalities, annotated by experienced sonographers using ellipses to fit the fetal head. The images are taken across all three trimesters (16\% first, 70\% second, 14\% third). To fine-tune the SD model with LoRA, 20 images per trimester were randomly selected to form dataset \{\boxnumber{2}\,,\,\boxnumber{3}\} as shown in Figure~\ref{fig:our_methods}(a). For SAM fine-tuning, the dataset was split 70:30 into training and test sets, with 50 test samples used for validation.
The {\bf ES} dataset includes images from the second and third trimesters, excluding cases with multiple pregnancies, congenital malformations, or aneuploidies~\cite{Xavier:2020,Alzubaidi:2023}. These are annotated by students, physicians, and radiologists. We randomly selected 500 trans-thalamic images to fine-tune SAM and 597 trans-ventricular images for evaluation.
The {\bf AF} dataset contains 125 images from five countries and was used as an out-of-distribution test set for our fine-tuned SAM $f_\text{SAM}$. Only images taken across the second and third trimesters are included.

\paragraph{Baseline Data Augmentation Methods:} The baseline data augmentation methods include two types: {\bf Weak Augmentation (WA)} and {\bf Strong Augmentation (SA)}. The weak augmentation techniques are horizontal flip, vertical flip, rotation, brightness and contrast, blur, and Gaussian noise. The strong augmentation techniques are rotation, color jitter, elastic deformation, and random erasing. The detailed parameters of these data augmentation techniques can be found within our code repository on GitHub. 


\subsection{Implementation Details}

\noindent {\bf LoRA Fine-tuning:} LoRA learning rate 1e-04, training epochs 1, batch size 4, LoRA rank 128, learning scheduler: constant. Stable Diffusion v1.5 is used as the base model for fine-tuning. We use UniPC sampler, LoRA weight of 0.9, and the number of sampling steps 20 for image generation. Three LoRA models are trained, one for each trimester. We have 20 real images from each trimester used for fine-tuning. Our implementation is done in PyTorch. LoRA Fine-tuning experiments are performed on a single NVIDIA A100 GPU. The threshold value utilized in the Ellipse Extractor is set to 127.

\noindent {\bf Fine-tuning SAM:} We evaluate the performance of our proposed augmentation technique on fetal head segmentation tasks. For the segmentation model, we use pre-trained SAM~\cite{kirillov:2023}. 
We fine-tune the mask decoder of the pre-trained SAM for 20 epochs, the labeled batch size is set to 5, the optimizer is Adam~\cite{Kingma:2015}, the learning rate is $1e-5$, and weight decay is $0$. Our code is developed in Python 3.11.5 using Pytorch 2.1.2 and CUDA 12.2 using one NVIDIA RTX 4090 GPU. 
The network is evaluated on the validation set every epoch, and the weight of the SAM is saved when the performance on validation outperforms the best previous performance.

\noindent {\bf Segmentation Evaluation Metrics:} The comparative experiments between our augmentation method and other baseline data augmentation methods are carried out, utilizing evaluation metrics encompassing a similarity measure: Dice Score (DSC). We run tests for each case 5 times repeatedly; see Table \ref{overall_res}.

\noindent {\bf Loss Function of SAM:} For fine-tuning SAM, the loss function is the sum of Dice and Cross Entropy losses, which can be expressed as:

\begin{equation}
\label{sam_loss}
    \mathcal{L}_{\text {SAM}}= \mathcal{L}_\text{R} + \mathcal{L}_\text{S} = \underbrace{\operatorname{CE}\left(\hat{\boldsymbol Y}_\text{R}, \bm Y_\text{R}\right) + 
    \operatorname{Dice}\left(\hat{\boldsymbol Y}_\text{R}, \bm Y_\text{R}\right)}_\text{Real Data} +
    \underbrace{\operatorname{CE}\left(\hat{\boldsymbol Y}_\text{S}, \bm Y_\text{S}\right) + 
    \operatorname{Dice}\left(\hat{\boldsymbol Y}_\text{S}, \bm Y_\text{S}\right)}_\text{Synthetic Data}
\end{equation}
where $\mathcal{L}_\text{R}$ and $\mathcal{L}_\text{S}$ represent the segmentation losses for $\boldsymbol{X}_\text{R}$ and $\boldsymbol{X}_\text{S}$, respectively. The segmentation predictions of real ultrasound images are $\hat{\boldsymbol Y}_\text{R} = f_\text{SAM}\left(\boldsymbol{X}_\text{R}\right)$ and synthetic images are $\hat{\boldsymbol Y}_\text{S} = f_\text{SAM}\left(\boldsymbol{X}_\text{S}\right)$. $\operatorname{CE}$ represents cross-entropy loss and $\operatorname{Dice}$ represents Dice loss.

\section{Results}
\label{res}

\begin{figure}
    \centering
    \includegraphics[width=.9\linewidth]{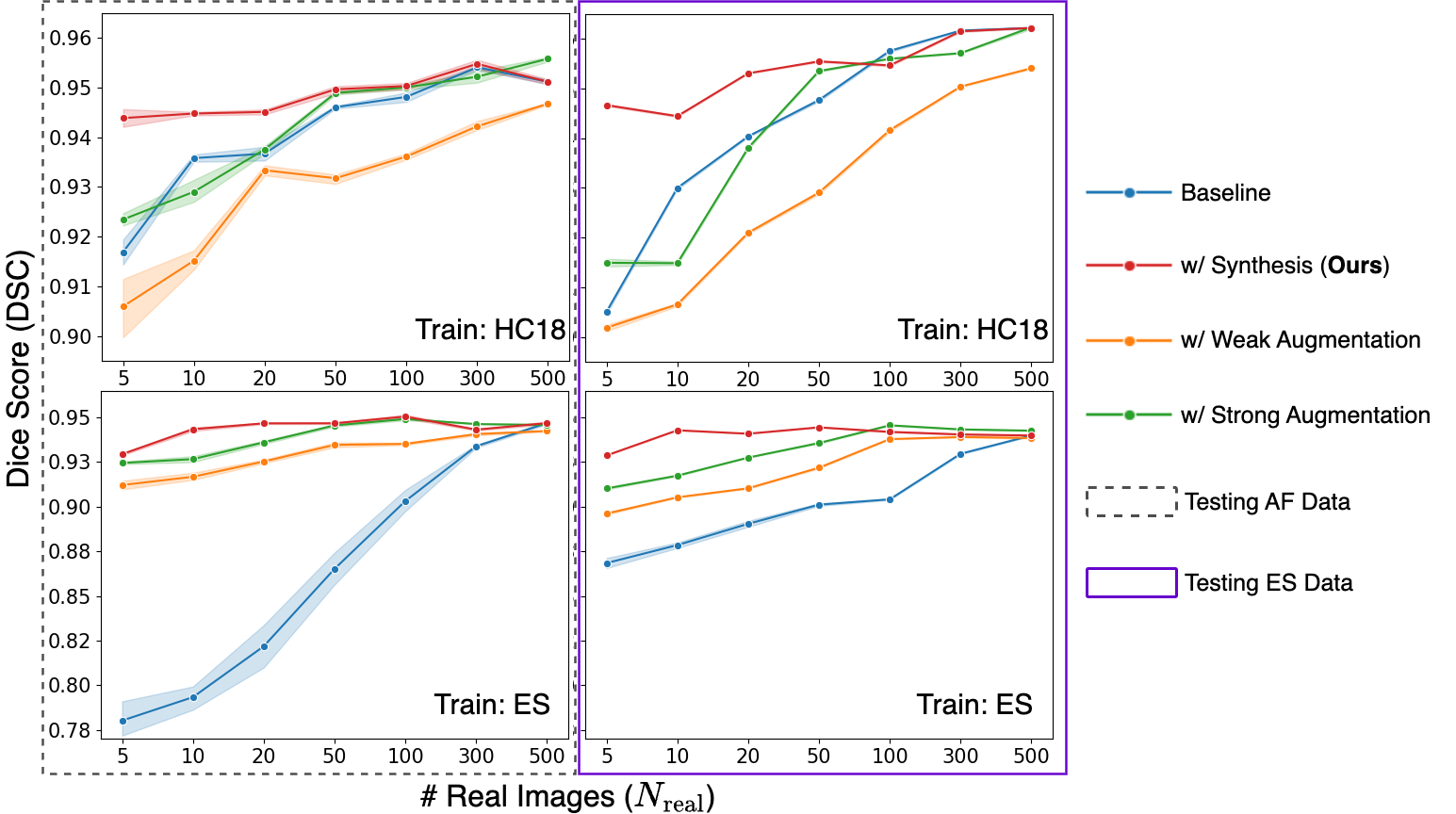}
    \caption{The plot shows the mean Dice score and standard deviation of fetal head segmentation using various augmentation techniques during training.}
    \label{fig:linechart}
    \vspace{-10pt}
\end{figure}

We conducted a comparative analysis of our method with SA and WA to illustrate its superiority in downstream segmentation tasks. An advantage of the GenAI method lies in its ability to augment the training dataset. Consequently, our method maintains a fixed training dataset size of 500 images. Specifically, the number of synthetic images generated is equal to $500 - N_{\text{real}}$. Both SA and WA have a training dataset size equal to $N_{\text{real}}$ as indicated in Table \ref{overall_res}.
In terms of the segmentation performance, Figure \ref{fig:linechart} demonstrates a significant improvement when fine-tuning SAM using synthetic image-mask pairs generated by our method, particularly with a small number of annotated training images. 
Furthermore, it is notable that our method results in a more robust segmentation performance with significantly lower standard deviation in the DSC when compared to other data augmentation techniques.
When comparing the performance across AF and ES datasets, our method produces a more generalized segmentation model with minimal performance disparities compared to other data augmentation methods.

\begin{wrapfigure}{r}{0.5\textwidth}
  \vspace{-28pt}
  \begin{center}
    \includegraphics[width=0.5\textwidth, height=0.3\textwidth]{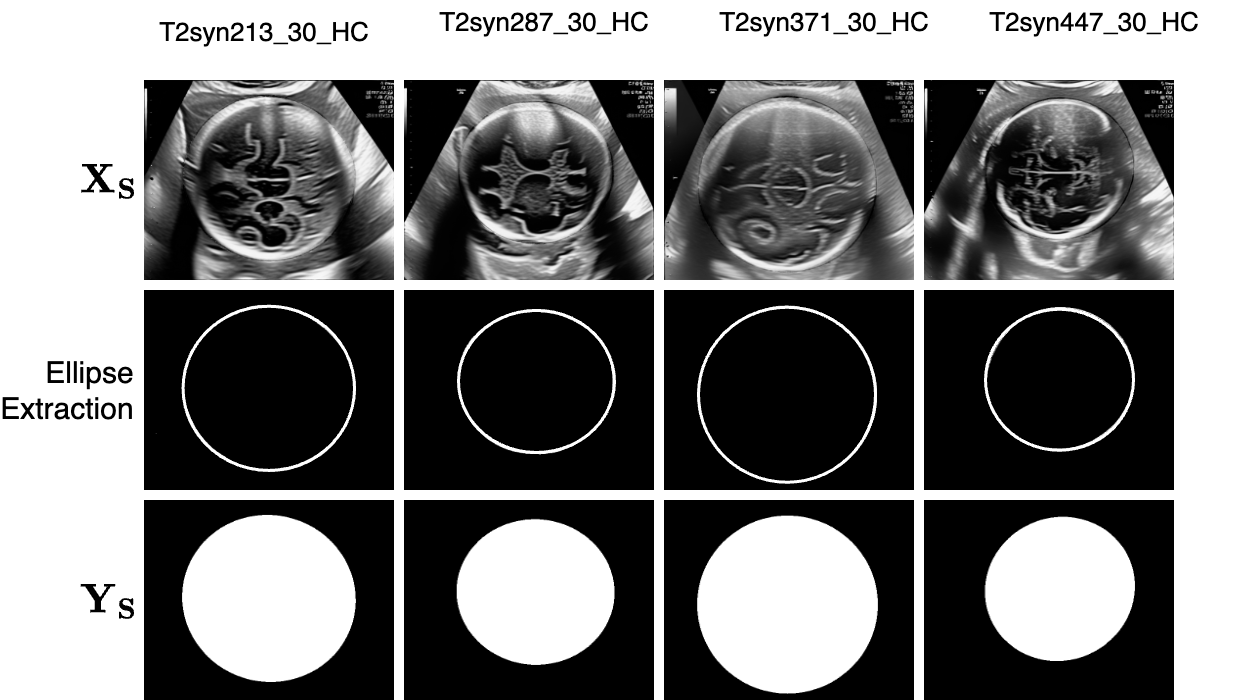}
\end{center}
\vspace{-10pt}
  \hspace*{0.13\linewidth}\parbox{0.84\linewidth}{\caption{The synthetic images $\boldsymbol{X}_\text{S}$ we generated and their ground truth masks.}\label{fig:vis_samples}}
  \vspace{-10pt}
\end{wrapfigure}

Figure \ref{fig:vis_samples} displays a selection of synthetic fetal ultrasound images generated by our fine-tuned SD model. The skull is prominently visible in the synthetic brain images $\boldsymbol{X}_\text{S}$ as a bright, elliptical structure, reproducing the characteristics observed in real-world images. The ellipse extraction shows that the simple thresholding method can precisely extract the shape of the fetal skull as a binary mask. After filling the ellipse shape, we obtain the ground truth mask $\boldsymbol{Y}_\text{S}$ for each $\boldsymbol{X}_\text{S}$.
\vspace{0.3em}

\begin{table}[htbp]
\centering
\caption{The segmentation results when 5, 10, 20, 50, 100, 300, and 500 real images for training. The best results are in {\bf bold}. DSC: Dice Score (\%). WA: Weak augmentation. SA: Strong augmentation. SYN: Synthetic data augmentation.}
\label{overall_res}
\begin{tabular}{|l|l|ccccccc|}
\hline
\multirow[c]{2}{*}{\bf DA} & \multirow[c]{2}{*}{\bf Train/Test} & \multicolumn{7}{c|}{\bf {\# Real Images ($N_\text{real}$)} -- ({mean DSC $\uparrow$ $\pm$ std $\downarrow$})} \\
 &  & \bf 5 & \bf 10 & \bf 20 & \bf 50 & \bf 100 & \bf 300 & \bf 500 \\
\hline\hline

- (baseline) & \multirow{4}{*}{HC18 / ES} & 90.51$\pm$0.05 & 93.00$\pm$00.04 & 94.03$\pm$0.02 & 94.76$\pm$0.04 & 95.75$\pm$0.03 & \bf 96.16$\pm$0.02 & 96.22$\pm$0.01 \\
WA &                             & 90.19$\pm$0.09 & 90.66$\pm$0.04 & 92.10$\pm$0.04 & 92.90$\pm$0.04 & 94.16$\pm$0.04 & 95.04$\pm$0.02 & - \\ 
SA &                             & 91.49$\pm$0.10 & 91.49$\pm$0.05 & 93.80$\pm$0.04 & 95.35$\pm$0.02 & \bf 95.60$\pm$0.01 & 95.71$\pm$0.01 & - \\ 
SYN (Ours) &                   & \bf 94.66$\pm$0.03 & \bf 94.44$\pm$0.02 & \bf 95.30$\pm$0.01 & \bf 95.54$\pm$0.02 & 95.46$\pm$0.02 & 96.15$\pm$0.02 & - \\       
\hline            
- (baseline) & \multirow{4}{*}{HC18 / AF} & 91.69$\pm$0.34 & 93.58$\pm$0.09 & 93.67$\pm$0.18 & 94.60$\pm$0.04 & 94.81$\pm$0.12 & 95.40$\pm$0.06 & 95.11$\pm$0.06 \\
WA &                             & 90.60$\pm$0.76 & 91.52$\pm$0.25 & 93.34$\pm$0.13 & 93.17$\pm$0.13 & 93.61$\pm$0.08 & 94.22$\pm$0.12 & - \\ 
SA &                             & 92.35$\pm$0.16 & 92.90$\pm$0.27 & 93.75$\pm$0.10 & 94.89$\pm$0.05 & 95.00$\pm$0.05 & 95.22$\pm$0.16 & - \\ 
SYN (Ours) &                   & \bf 94.38$\pm$0.23 & \bf 94.48$\pm$0.04 & \bf 94.51$\pm$0.07 & \bf 94.96$\pm$0.09 & \bf 95.03$\pm$0.08 & \bf 95.47$\pm$0.12 & - \\
\hline\hline
- (baseline) & \multirow{4}{*}{ES / ES} & 86.84$\pm$0.37 & 87.85$\pm$0.22 & 89.04$\pm$0.23 & 90.11$\pm$0.10 & 90.41$\pm$0.05 & 92.96$\pm$0.05 & 93.99$\pm$0.03 \\
WA &                             & 89.62$\pm$0.07 & 90.53$\pm$0.03 & 91.03$\pm$0.04 & 92.18$\pm$0.06 & 93.79$\pm$0.03 & 93.90$\pm$0.03 & - \\ 
SA &                             & 91.02$\pm$0.07 & 91.74$\pm$0.05 & 92.75$\pm$0.02 & 93.57$\pm$0.05 & \bf 94.56$\pm$0.04 & \bf 94.33$\pm$0.03 & - \\ 
SYN (Ours) &                   & \bf 92.87$\pm$0.03 & \bf 94.28$\pm$0.02 & \bf 94.09$\pm$0.01 & \bf 94.44$\pm$0.02 & 94.19$\pm$0.04 & 94.05$\pm$0.04 & - \\
\hline                    
- (baseline) & \multirow{4}{*}{ES / AF} & 78.02$\pm$1.20 & 79.33$\pm$0.90 & 82.18$\pm$1.51 & 86.52$\pm$1.23 & 90.32$\pm$0.78 & 93.38$\pm$0.11 & 94.68$\pm$0.09 \\
WA &                             & 91.21$\pm$0.32 & 91.67$\pm$0.25 & 92.53$\pm$0.14 & 93.46$\pm$0.17 & 93.52$\pm$0.08 & 94.06$\pm$0.10 & - \\ 
SA &                             & 92.45$\pm$0.10 & 92.66$\pm$0.22 & 93.60$\pm$0.12 & 94.55$\pm$0.09 & 94.91$\pm$0.08 & \bf 94.63$\pm$0.01 & - \\ 
SYN (Ours) &                   & \bf 92.94$\pm$0.09 & \bf 94.34$\pm$0.13 & \bf 94.67$\pm$0.05 & \bf 94.67$\pm$0.05 & \bf 95.05$\pm$0.14 & 94.31$\pm$0.07 & - \\                     
\hline
\end{tabular}
\end{table}

Table \ref{overall_res} shows the quantitative results of fetal head segmentation performance with SAM. When the number of annotated real images is fewer than 50, our diffusion-based data augmentation method improves SAM's segmentation performance to achieve the SOTA results. Significantly, our method outperforms the baseline in the ES and AF datasets when using 10 and 100 real ES images for training. Although the gaps between our method and other techniques lessen with the expansion of the real dataset for training, we still achieve superior overall performance through a few-shot learning strategy in fetal head segmentation. 

\section{Conclusion}


In this study, we present a novel model-based data augmentation method with a fine-tuned diffusion model to enhance the efficiency and effectiveness of fetal head segmentation in ultrasound imaging. Our study demonstrates that the fine-tuned text-to-image diffusion model can produce a coherent synthetic image along with a segmentation mask from a given text prompt. Moreover, our proposed approach requires only 20 fetal ultrasound images from each trimester for fine-tuning, allowing the generation of an extensive set of synthetic image-mask pairs to augment training data for segmentation tasks. By constructing a hybrid dataset to fine-tune SAM, we have achieved SOTA Dice scores compared to various data augmentation techniques, particularly in scenarios with a limited number of real image-mask pairs (less than 50).
In conclusion, our innovative mask-guided diffusion model approach leads to time and effort savings in the analysis of fetal ultrasound imaging compared to manual annotation.
For future work, we plan to extend our method to segment additional fetal anatomical structures and explore cross-domain generalization across different populations and ultrasound devices. Furthermore, we aim to integrate temporal consistency into the diffusion-based generation process to enable synthetic video-based ultrasound data generation.

\section*{Acknowledgments} 
This work was funded by Taighde \'{E}ireann – Research Ireland through the Research Ireland Centre for Research Training in Machine Learning (18/CRT/6183).







\bibliographystyle{apalike}
\small
\bibliography{imvip}

\end{document}